\documentclass[prd,aps,floats,floatfix,nofootinbib]{revtex4}
\usepackage{amsmath}
\usepackage{amsfonts}
\usepackage{graphicx}
\usepackage{slashed}
\usepackage{dcolumn}
\usepackage{bm}
\usepackage[dvips]{color}
\newcommand{\beq}{\begin{equation}}
\newcommand{\eeq}{\end{equation}}
\begin{document}
\newcommand{\vect}[1]{\overrightarrow{#1}}
\newcommand{\smbox}[1]{\mbox{\scriptsize #1}}
\newcommand{\tanbox}[1]{\mbox{\tiny #1}}
\newcommand{\vev}[1]{\langle #1 \rangle}
\newcommand{\Tr}[1]{\mbox{Tr}\left[#1\right]}
\newcommand{\cosb}{c_{\beta}}
\newcommand{\sinb}{s_{\beta}}
\newcommand{\tanb}{t_{\beta}}
\newcommand{\picwidth}{3.4in}

\preprint{MSU-HEP-100201}
\title{Custodial Isospin Violation in the Lee-Wick Standard Model}

\author{R. Sekhar Chivukula}
\email[]{sekhar@msu.edu}
\author{Arsham Farzinnia}
\email[]{farzinni@msu.edu}
\author{Roshan Foadi}
\email[]{foadiros@msu.edu}
\author{Elizabeth H. Simmons}
\email[]{esimmons@msu.edu}
\affiliation{Department of Physics,
Michigan State University, East Lansing, MI 48824, USA}
\date{\today}

\begin{abstract}
We analyze the tension between naturalness and isospin violation in the Lee-Wick Standard Model (LW SM), by computing tree-level and fermionic one-loop contributions to the post-LEP electroweak parameters ($\hat{S}$, $\hat{T}$, $W$, and $Y$) and the $Z b_L \bar{b}_L$ coupling.  The model is most natural when the LW partners of the gauge bosons and fermions are light, but small partner masses can lead to large isospin violation. The post-LEP parameters yield a simple picture in the LW SM:  the gauge sector contributes to $Y$ and $W$ only, with leading contributions arising at tree-level, while the fermion sector contributes to $\hat{S}$ and $\hat{T}$ only, with leading corrections arising at one loop.  Hence, $W$ and $Y$ constrain the masses of the LW gauge bosons to satisfy  $M_1, M_2  \gtrsim 2.4$ TeV at 95\% CL.  Likewise, experimental limits on $\hat{T}$ reveal that the masses of the LW fermions must satisfy $M_q, M_t  \gtrsim 1.6$ TeV at 95\% CL if the Higgs mass is light and tend to exclude the LW SM for any LW fermion masses if the Higgs mass is heavy.  Contributions from the top-quark sector to the $Zb_L \bar{b}_L$ coupling can be even more stringent, placing a lower bound of 4 TeV on the LW fermion masses at 95\% CL.  

\end{abstract}

\maketitle

\section{Introduction}

Recently an extension of the Standard Model (SM) based on the ideas of Lee and Wick~\cite{Lee:1969fy,Lee:1970iw} has been proposed\footnote{A Lee-Wick extension of the Higgs sector had been previously proposed in~\cite{Jansen:1992xv,Jansen:1992xx,Jansen:1993ji,Jansen:1993jj}.}~\cite{Grinstein:2007mp}. This features higher derivative kinetic terms for each SM field. As a consequence the field propagators fall off to zero more rapidly than the ordinary SM propagators, and the infinities associated with ultraviolet quantum fluctuations are either softened or even removed from the theory. In a scalar field theory all amplitudes turn out to be finite by power counting. In a gauge theory the higher derivative kinetic terms generate new momentum-dependent interactions which prevent the theory from being finite; however, a simple power counting argument shows that the only possible divergences are logarithmic. Thus the LW SM offers a potential solution to the hierarchy problem. This was the main motivation for studying the model in~\cite{Grinstein:2007mp}, and analyzing its phenomenological implications~\cite{Rizzo:2007ae,Rizzo:2007nf,Alvarez:2009af,Krauss:2007bz}.

If a higher-derivative kinetic term is added to the Lagrangian, the propagator of a LW SM field displays two poles, the lighter one corresponding to the ordinary SM particle, and the heavier one corresponding to a new degree of freedom, the LW partner. An equivalent formulation consists of separating the poles, in such a way that to each field there corresponds only one pole and one mass. The LW poles are then characterized by a negative residue, and thus act as Pauli-Villar regulators. However, unlike mere regulators, the LW fields nontrivially participate in gauge and Yukawa interactions.

In the LW SM, as in the ordinary SM, the largest one-loop contribution to the Higgs mass comes from an isospin violating sector of the theory: the top Yukawa coupling. There are two heavy partners of the top quark in the LW SM, one associated with the left-handed top-bottom doublet, with mass $M_q$, and the other with the right-handed top, with mass $M_t$. The contributions to the Higgs mass involving a single LW top are opposite in sign to those from a single SM top, so they cancel the quadratic divergence in $\delta m_h^2$. The net contribution is still logarithmically divergent, and for degenerate LW top quarks, $M_q=M_t$, is of the form
\begin{equation}
\delta m_h^2=\frac{3\lambda_t^2}{8\pi^2}M_q^2 \log\frac{\Lambda^2}{M_q^2} \ ,
\end{equation} 
where $\Lambda$ is the cutoff. In the limit $M_q\to\infty$  the ordinary quadratic divergence reappears, with $M_q$ acting as a cutoff. Therefore, as already pointed out in Ref.~\cite{Alvarez:2008ks}, in order to avoid fine tuning  the value of $M_q$ cannot be too large. 

Because the dominant correction to the Higgs mass is associated with an isospin violating sector of the theory, it is important to check whether the LW tops cause a large contribution to the electroweak observables which are usually protected by custodial symmetry: $\Delta\rho$, and, for theories with heavy replicas of the top quark, the $Z b_L \bar{b}_L$ coupling~\cite{Agashe:2006at}~\cite{SekharChivukula:2009if}. Large contributions to these quantities would lead to a stringent lower bound on $M_q$, which would result in large corrections to $m_h^2$ and thus the necessity of fine-tuning the scalar sector of the theory.

In this paper we analyze the potential conflict between naturalness and isospin violation, by computing the contribution of the top-quark sector to the $\rho$ parameter and to the $Z b_L \bar{b}_L$ coupling. Furthermore, we compute the Barbieri {\em et. al.}  ~\cite{Barbieri:2004qk,Chivukula:2004af} post-LEP electroweak parameters ($\hat{S}$, $\hat{T}$, $W$, and $Y$) to check for additional constraints. In terms of the post-LEP parameters, 
we find a simple picture for the constraints on the Lee-Wick standard model.  The dominant contributions to $\hat{T}$ come from the fermion sector at one loop, and limits on this parameter provide the strongest constraints on the top-quark sector\footnote{The dominant contributions to $\hat{S}$ likewise come from the fermion sector at one loop, but they are too small to provide strong constraints on the top quark sector.} .  In contrast, the dominant contributions
to $Y$ and $W$ arise from the gauge sector at tree-level, and limits on these parameters therefore
provide the strongest constraints on the gauge sector. These results imply that the bounds on the LW fermions, coming almost entirely from $\hat{T}$, are essentially independent of the LW gauge masses.

Our results differ from those in Refs.~\cite{Alvarez:2008ks, Carone:2008bs} because their one-loop analysis of the effects of LW top quarks on electroweak observables rests on the incorrect assumption that the corrections are purely oblique \cite{Peskin:1991sw,Altarelli:1990zd,Altarelli:1991fk}.  As discussed in Ref.~\cite{Underwood:2008cr} important non-oblique corrections arise at tree-level in the LW SM, in the form of non-zero values for $W$ and $Y$.  Therefore, one must use the Barbieri {\em et. al.}  parameters to compare the LW SM with experiment.

In section~\ref{sec:model}, we review the structure of the Lee-Wick standard model~\cite{Grinstein:2007mp}
and establish notation.  In section~\ref{sec:effective} we present an effective field theory analysis of
the LW corrections to $\Delta \rho$ and to the $Zb_L\bar{b}_L$ coupling. 
In section~\ref{sec:parameters} we present our analysis of the post-LEP electroweak parameters and the resulting constraints on the Lee-Wick standard model, while the constraints from the $Z b_L \bar{b}_L$ coupling are analyzed in section~\ref{sec:Zbb}. The leading logarithmic contributions to the electroweak observables in the full theory and the effective theory have to match, thus the results of section~\ref{sec:effective} provide an important check for those of sections~\ref{sec:parameters} and \ref{sec:Zbb}. 

Questions concerning unitarity~\cite{Cutkosky:1969fq}, causality~\cite{Grinstein:2008bg}, and Lorentz invariance in LW theories, although potentially important, will not be considered in this analysis. A complete analysis of one-loop renormalization of the LW SM can be found in~\cite{Grinstein:2008qq}.


\section{The Lee-Wick Standard Model}\label{sec:model}

It is straighforward to write a higher derivative extension of the SM electroweak Lagrangian
\cite{Grinstein:2007mp}. Adopting a non-canonical normalization for the gauge fields, the gauge Lagrangian reads
\begin{eqnarray}
{\cal L}_{\rm gauge}^{\rm hd} = -\frac{1}{4g_1^2}\hat{B}_{\mu\nu}\hat{B}^{\mu\nu}
-\frac{1}{2 g_2^2}{\rm Tr}\left[\hat{W}_{\mu\nu}\hat{W}^{\mu\nu}\right]
+\frac{1}{2 g_1^2 M_1^2}\partial^\mu \hat{B}_{\mu\nu} \partial_\lambda \hat{B}^{\lambda\nu}
+\frac{1}{g_2^2 M_2^2}{\rm Tr}\left[\hat{D}^\mu \hat{W}_{\mu\nu} \hat{D}_\lambda \hat{W}^{\lambda\nu}\right] \ .
\label{eq:gaugelagrg}
\end{eqnarray}
The ``hat" notation indicates that the field's propagator contains not only the ordinary SM poles but also the LW poles. For example, in the limit of unbroken electroweak phase the $\hat{B}_\mu$ propagator has a massless pole, corresponding to the ordinary $B_\mu$ gauge field, and a mass-$M_1$ pole, corresponding to its LW counterpart. Notice also that additional dimension-six operators, quadratic in the field strength tensors could, in principle, be added to this Lagrangian. However these would lead to scattering amplitudes for the longitudinally polarized gauge bosons growing like $E^2$, and thus to a rather early violation of unitarity~\cite{Grinstein:2007iz}. We therefore do not include them in this analysis. Notice also that we only include one higher derivative term per SM field, which introduces a single corresponding LW pole. This is certainly fine for our purposes, since in this analysis we focus on the low momentum regime, where additional higher derivative terms are negligible. However at large momenta additional poles in the propagator can have important implications~\cite{Carone:2008iw,Carone:2009it}.
 
The higher derivative extension of the Higgs sector is 
\begin{eqnarray}
{\cal L}_{\rm Higgs}^{\rm hd} = |\hat{D}_\mu \hat{\phi}|^2
-\lambda\left(\hat{\phi}^\dagger \hat{\phi} -\frac{v^2}{2}\right)^2 
-\frac{1}{M_h^2} |\hat{D}^2 \hat{\phi}|^2 \ ,
\label{eq:higgslagrg}
\end{eqnarray}
where as usual the Higgs doublet may be written in component form as
\begin{eqnarray}
\hat{\phi} = \frac{1}{\sqrt{2}}
\left(\begin{array}{c}
i\sqrt{2}\hat{\phi}^+ \\ v+\hat{h}-i\hat{\phi}^0
\end{array}
\right)~.
\end{eqnarray}
Here and in Eq.~(\ref{eq:gaugelagrg}) the covariant derivative written with a hat is built with the hatted gauge fields. We will find it convenient to have a compact way of denoting $i \sigma^2 \hat{\phi}^\ast$ as we build operators that couple the Higgs to the right-handed top quark.  Hence, we make the definition:
\begin{eqnarray}
\hat{\varphi} \equiv (i \sigma^2 \hat{\phi}^\ast) =  
\frac{1}{\sqrt{2}}
\left(\begin{array}{c}
v+\hat{h}+i\hat{\phi}^0\\
i\sqrt{2}\hat{\phi}^- 
\end{array}\right)~.
\end{eqnarray}
The field $\hat{\phi}$ contains both the ordinary Higgs doublet and a massive doublet\footnote{If $M_h$ is smaller than all other LW mass parameters, in a certain energy regime the model behaves like a two-Higgs doublet model, although one doublet is of LW type. This scenario was analyzed in~\cite{Carone:2009nu}.}
 with mass $\sim M_h$. 

In the fermion sector we only focus on the third quark generation, since this is the dominant source of isospin violation, and gives the largest correction to the Higgs mass.\footnote{Inclusion of the remaining flavors would introduce new mixing matrices, and, without the assumption of minimal flavor violation, potential sources of flavor changing neutral current. For a discussion see Ref.~\protect\cite{Dulaney:2007dx}.} The higher-derivative extension of the fermion Lagrangian is
\begin{eqnarray}
{\cal L}_{\rm quark}^{\rm hd} = \bar{\hat{q}}_L i \hat{\slashed{D}} \hat{q}_L
+ \bar{\hat{t}}^\prime_R  i \hat{\slashed{D}} \hat{t}^\prime_R
+ \bar{\hat{b}}^\prime_R  i \hat{\slashed{D}} \hat{b}^\prime_R
+\frac{1}{M_q^2} \bar{\hat{q}}_L  i \hat{\slashed{D}}^3 \hat{q}_L
+\frac{1}{M_t^2} \bar{\hat{t}}^\prime_R  i \hat{\slashed{D}}^3 \hat{t}^\prime_R
+\frac{1}{M_b^2} \bar{\hat{b}}^\prime_R  i \hat{\slashed{D}}^3 \hat{b}^\prime_R \ ,
\label{eq:higher}
\end{eqnarray}
where $\hat{q}_L=(\hat{t}_L,\hat{b}_L)$. Notice that the right handed fields have been primed because, for example, $\hat{t}_L$ and $\hat{t}^\prime_R$ are not left and right component of the same Dirac spinor. In the unbroken electroweak phase $\hat{t}_L$ ($\hat{t}^\prime_R$) contains the ordinary massless SM left-handed (right-handed) top as well as a massive Dirac fermion of mass $M_q$ ($M_t$). 

Finally we consider the Yukawa Lagrangian, which in the LW SM has no derivatives. Therefore, we write
\begin{eqnarray}
{\cal L}_{\rm Yukawa} = -y_t\ \bar{\hat{q}}_L\  \hat{\varphi}\ \hat{t}^\prime_R \ +\ {\rm h.c.} \ ,
\label{eq:yukhigher}
\end{eqnarray}
where the bottom Yukawa coupling has been ignored, since $y_b\ll y_t$.

As explained in the introduction this ``higher derivative" formulation of the theory, in which both the ordinary pole and the LW pole are contained in the same field,  is equivalent to an ``ordinary formulation" in which: (i) the two poles belong to two different fields, and (ii) the kinetic and mass terms for the LW fields have the ``wrong'' sign. This alternative formulation is especially useful for calculating loop diagrams. In this paper we will compute loop diagrams with the top and bottom quarks in the loop. Thus we will find it helpful to replace the higher derivative fermion and Yukawa Lagrangians with the ordinary formulation Lagrangians
\begin{eqnarray}
{\cal L}_{\rm quark} = \bar{q}_L i \hat{\slashed{D}} q_L
+ \bar{t}^\prime_R i \hat{\slashed{D}} t^\prime_R
+ \bar{b}^\prime_R i \hat{\slashed{D}} b^\prime_R
-\bar{\tilde{q}}\left(i\hat{\slashed{D}}-M_q\right)\tilde{q}
-\bar{\tilde{t}}^\prime\left(i\hat{\slashed{D}}-M_t\right)\tilde{t}^\prime
-\bar{\tilde{b}}^\prime\left(i\hat{\slashed{D}}-M_b\right)\tilde{b}^\prime \ ,
\label{eq:lower}
\end{eqnarray}
and
\begin{eqnarray}
{\cal L}_{\rm Yukawa} = -y_t\ \left(\bar{q}_L-\bar{\tilde{q}}_L\right)\ \hat{\varphi}\ 
\left(t^\prime_R-\tilde{t}^\prime_R\right) \ + \ {\rm h.c.} \ ,
\label{eq:yuklower}
\end{eqnarray}
where
\begin{eqnarray}
\hat{q}_L \equiv q_L - \tilde{q}_L\ , \quad\quad \hat{t}^\prime_R\equiv t^\prime_R-\tilde{t}^\prime_R\ , \quad\quad 
\hat{b}^\prime_R\equiv b^\prime_R-\tilde{b}^\prime_R \ ,
\end{eqnarray}
and where the fields with (without) a tilde are LW (SM) fields.
The equivalence between the higher derivative formulation, Eqs.~(\ref{eq:higher},\ref{eq:yukhigher}), and the ordinary formulation, Eqs.~(\ref{eq:lower},\ref{eq:yuklower}), can be easily proved; see for example~\cite{Grinstein:2007mp}. Notice that the ``wrong" sign in front of the kinetic and mass term makes the LW (tilde) fields act like Pauli-Villars regulators, with the difference that they also participate nontrivially in gauge and Yukawa interactions.


\section{Effective Field Theory for $\Delta \rho$ and $Zb\bar{b}$}\label{sec:effective}

The appearance of the LW fields in the Yukawa interactions, Eq.~(\ref{eq:yuklower}), suggest the presence of nonstandard sources of custodial isospin violation at energies below the LW scale. Dimension-six custodial violating operators can potentially arise from tree-level exchanges, and from loop diagrams with one or more LW fermions in the loop. The leading contribution to these operators, in inverse powers of the LW fermion masses, can be found by integrating out the LW fermions at tree-level, and computing loops in the resulting effective field theory. For LW fermion masses much larger than both the Higgs vacuum and the external momenta, the effective Lagrangian can be computed in powers of $\hat{\varphi}/M_{q,t}$ and $\hat{\slashed{D}}/M_{q,t}$. Including the leading nonstandard corrections, this leads to
\begin{eqnarray}
{\cal L}_{\rm eff} &=& \bar{q}_L i \hat{\slashed{D}} q_L
+ \bar{t}_R i \hat{\slashed{D}} t_R
+ \bar{b}_R i \hat{\slashed{D}} b_R
- y_t\left( \bar{q}_L \hat{\varphi}\ t_R + \bar{t}_R \hat{\varphi}^\dagger q_L \right) \nonumber \\
&-& \frac{y_t^2}{M_t^2}\ \bar{q}_L i \hat{\varphi}\hat{\slashed{D}}\left(\hat{\varphi}^\dagger q_L \right)
- \frac{y_t^2}{M_q^2}\ \bar{t}_R  i \hat{\varphi}^\dagger \hat{\slashed{D}}\left(\hat{\varphi} t_R \right) \ .
\label{eq:efflagr0}
\end{eqnarray}
Notice that the primes have been removed from the right-handed fermion fields, because now left-handed and right-handed components are Dirac partners. Notice also that this Lagrangian assumes $M_q$ and $M_t$ to be of the same order, with no hierarchy between them. The leading logarithmic correction to observables will therefore be proportional to $\log M_q^2/v^2\sim \log M_t^2/v^2$. In what follows, we compute these leading-log corrections by constructing the operators which arise in
the effective theory appropriate for energy scales below $M_t\simeq M_q$, in which the LW partners have
been ``integrated out" but the top quark remains in the spectrum.

After electroweak symmetry breaking, the higher derivative operators lead to a renormalization of the fermion kinetic terms. An alternative approach consists of redefining $q_L$ and $t_R$ to make their kinetic terms canonically normalized both in the broken and the unbroken electroweak phase. This is achieved by the replacements
\begin{eqnarray}
q_L  & \to & \left[1+\frac{y_t^2}{2 M_t^2} \hat{\varphi}\hat{\varphi}^\dagger 
+ {\cal O}(1/M_t^3)\right] q_L \ , \nonumber \\
t_R  & \to & \left[1+\frac{y_t^2}{2 M_q^2} \hat{\varphi}^\dagger\hat{\varphi} 
+ {\cal O}(1/M_q^3)\right] t_R \ ,
\end{eqnarray}
which leads to a new Lagrangian, equivalent to ${\cal L}_{\rm eff}$:
\begin{eqnarray}
{\cal L}^\prime_{\rm eff} &=& \bar{q}_L i \hat{\slashed{D}} q_L
+ \bar{t}_R i \hat{\slashed{D}} t_R
+ \bar{b}_R i \hat{\slashed{D}} b_R \nonumber \\
&-& y_t\ \bar{q}_L \hat{\varphi}\left[1+\frac{y_t^2}{2}\left(\frac{1}{M_q^2}+\frac{1}{M_t^2}\right)
\hat{\varphi}^\dagger \hat{\varphi}\right]t_R + {\rm h.c.} \nonumber \\
&+& \frac{y_t^2}{2M_t^2}\ \bar{q}_L i \left[(\hat{D}_\mu\hat{\varphi})\hat{\varphi}^\dagger
-\hat{\varphi}(\hat{D}_\mu\hat{\varphi})^\dagger\right] \gamma^\mu q_L
+ \frac{y_t^2}{2M_q^2}\ \bar{t}_R  \gamma^\mu t_R\ i \left[(\hat{D}_\mu\hat{\varphi})^\dagger \hat{\varphi}
-\hat{\varphi}^\dagger(\hat{D}_\mu\hat{\varphi})\right] \ .
\label{eq:efflagr}
\end{eqnarray}
As expected, there are custodial violating dimension-six operators. However at tree-level there is no non-standard contribution to $\Delta\rho$ or the $Z b_L \bar{b}_L$ coupling.

\begin{figure}[!t]
\includegraphics[width=5.5in,height=2.8in]{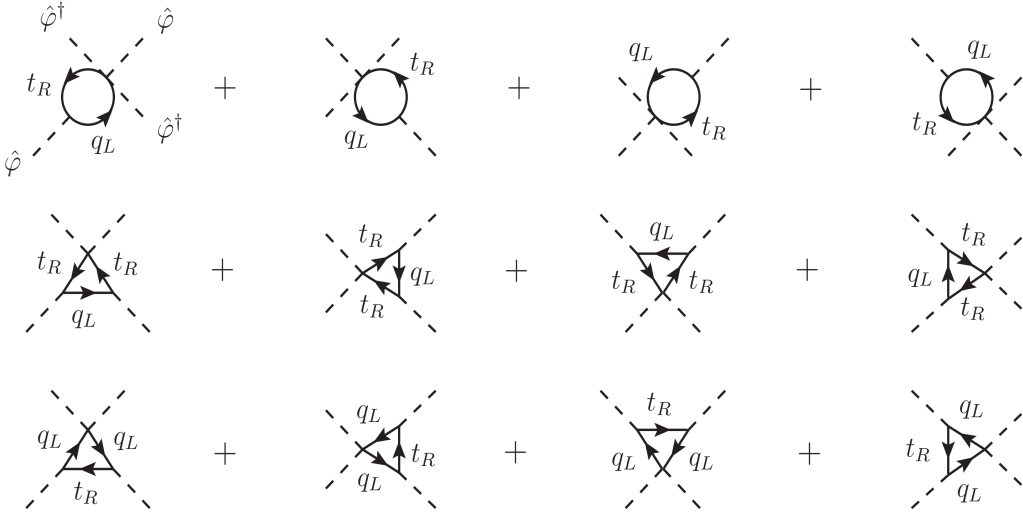}
\caption{Diagrams contributing to the dimension-six four-$\hat{\phi}$ operators in the effective theory, with the LW fermions integrated out at tree-level.}
\label{fig:effectiverho}
\end{figure}

${\cal L}^\prime_{\rm eff}$ features  terms coupling one, two, or three $\hat{\varphi}$ fields to a pair of fermions. Therefore, dimension-six four-$\hat{\varphi}$ operators arise both from vacuum polarization amplitudes and triangle diagrams, as shown in Fig.~\ref{fig:effectiverho}. The log-divergent parts of these diagrams (which yield the $\log(M^2_{t,q}/m^2_t)$ contributions)
 can be computed in the unbroken electroweak phase, with fermions in the loop. The logarithmically divergent part of the amplitude is reproduced by the operators\footnote{There are also quadratic divergences which are completely absorbed by a counterterm of the form $|\hat{\varphi}|^4$, with no derivatives.}
\begin{eqnarray}
\frac{3y_t^4}{16\pi^2}\left[\frac{2}{M_t^2}+\frac{1}{M_q^2}\right]|\hat{D}\hat{\phi}|^2 |\hat{\phi}|^2 \cdot \frac{1}{\epsilon}
+\frac{3y_t^4}{16\pi^2}\left[\frac{1}{M_t^2}+\frac{2}{M_q^2}\right]|\hat{\phi}^\dagger\hat{D}\hat{\phi}|^2 \cdot \frac{1}{\epsilon} \ ,
\end{eqnarray}
where as usual $\epsilon=2-d/2$ in dimensional regularization. The first operator respects custodial symmetry, but the second operator does not, since it only contributes to the $Z$ boson mass. The second operator gives the dominant contribution to $\Delta\rho$, which is therefore of the order
\begin{eqnarray}
(\Delta\rho)_{\rm LW} \sim -\frac{3}{16\pi^2}\frac{2m_t^4}{v^2}\left[\frac{1}{M_t^2}+\frac{2}{M_q^2}\right]\log{\frac{M_q^2}{m_t^2}} \ ,
\label{eq:rholeading}
\end{eqnarray}
where the $1/\epsilon$ is replaced by the large $\log$ which arises in the effective theory scaling from the scale $M_q\sim M_t$ to the weak scale $m_t\sim v$. For LW fermions lighter than 1 TeV this is a large {\it negative} isospin violating effect. For example, taking $M_q=M_t=500$ GeV gives $\Delta\rho\sim -1.4\%$. Furthermore, since $\Delta\rho$ is always negative, a heavy Higgs is strongly disfavored in the LW SM.

\begin{figure}
\includegraphics[width=5.5in,height=2.5in]{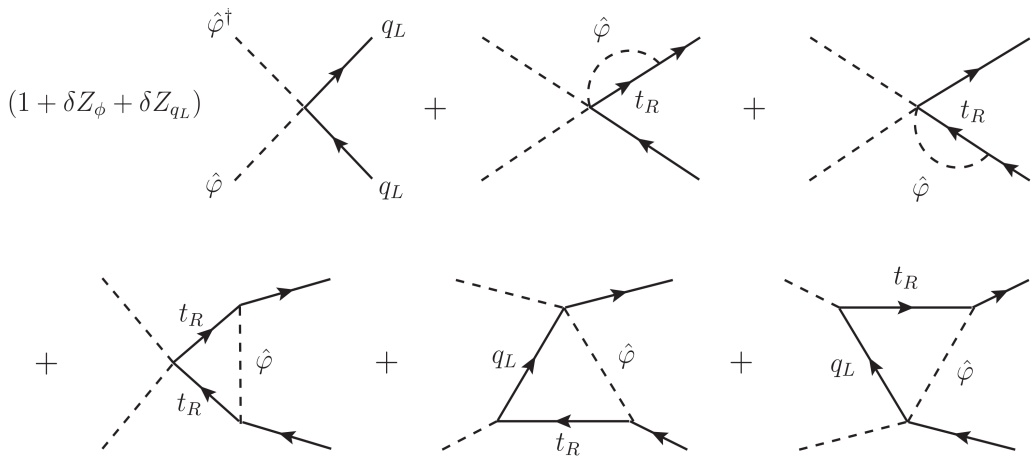}
\caption{Diagrams contributing to the dimension-six operators with two external $q_L$ and two $\hat{\varphi}$ fields. The triangle diagrams lead to the second operator of Eq.~(\ref{eq:Zbbeffective}), which contains non universal corrections to the $Z b_L \bar{b}_L$ coupling.}
\label{fig:effectiveZbb}
\end{figure}

The diagrams contributing to the  left-handed fermionic gauge couplings up to one-loop order are shown in Fig.~\ref{fig:effectiveZbb}. The tree-level diagram (corrected by the field strength renormalizations) corresponds to the custodial violating operator proportional to $y_t^2/2M_t^2$, in Eq.~(\ref{eq:efflagr}). This operator contributes to the $Z t_L \bar{t}_L$ coupling, not to $Z b_L \bar{b}_L$\footnote{Including the bottom Yukawa coupling would lead to a tree-level operator contributing to $Z b_L \bar{b}_L$. However the top-loop contribution is dominant, since $16\pi^2 y_b^2\sim 0.1$.}. The remaining diagrams contain nonstandard logarithmic divergences which are reproduced by the operators
\begin{eqnarray}
\frac{y_t^4}{16\pi^2}\frac{1}{4M_q^2} i \left[\bar{q}_L \gamma^\mu \hat{D}_\mu q_L - \bar{q}_L \overleftarrow{\hat{D}^\dagger_\mu} \gamma^\mu q_L\right]\hat{\varphi}^\dagger \hat{\varphi} \cdot \frac{1}{\epsilon}
+\frac{y_t^4}{16\pi^2}\left[\frac{1}{M_t^2}+\frac{1}{4 M_q^2}\right]
\ \bar{q}_L \gamma^\mu q_L \ i \left[(\hat{D}_\mu\hat{\varphi})^\dagger \hat{\varphi}
-\hat{\varphi}^\dagger(\hat{D}_\mu\hat{\varphi})\right]  \cdot \frac{1}{\epsilon} \ .
\label{eq:Zbbeffective}
\end{eqnarray}
In this expression the first (custodially symmetric) operator amounts to a renormalization of the standard gauge interactions, and does not contribute to nonstandard fermionic gauge couplings. The second operator violates custodial symmetry, and is only due to the triangle diagrams in Fig.~\ref{fig:effectiveZbb}. This contributes both to the $Z t_L \bar{t}_L$ coupling and the $Z b_L \bar{b}_L$ coupling. Expressing the latter in the form
\begin{equation}
\frac{e}{c_w s_w} g_L^{b\bar{b}}\ Z_\mu\  \bar{b}_L\gamma^\mu b_L \equiv \frac{e}{c_w s_w} \ \left[-\frac{1}{2}  + \frac{1}{3}\sin ^2 \theta_W 
+ (\delta g_L^{b\bar{b}})_{\rm SM} + (\delta g_L^{b\bar{b}})_{\rm LW} \right] \ Z_\mu\  \bar{b}_L\gamma^\mu b_L \ ,
\label{eq:totalZbb}
\end{equation}
where $(\delta g_L^{b\bar{b}})_{\rm SM}$ includes all higher order SM corrections, and replacing the $1/\epsilon$ poles with the large log arising from scaling in the
theory, we find that the second 
operator of Eq.~(\ref{eq:Zbbeffective}) gives the dominant non universal LW contribution to $g_L^{b\bar{b}}$:
\begin{eqnarray}
(\delta g_L^{b\bar{b}})_{\rm LW} \sim -\frac{m_t^4}{32\pi^2 v^2}\left[\frac{4}{M_t^2}+\frac{1}{M_q^2}\right]\log\frac{M_q^2}{m_t^2} \ .
\label{eq:Zbbleading}
\end{eqnarray}
The SM prediction is already 1.96$\sigma$ below the observed central value. Hence the additional 
 negative correction in the LW theory goes in the direction opposite 
to what is favored by experiment.

In the next two sections we compute perturbatively (in $v^2/M_q^2$ and $v^2/M_t^2$) and numerically
the values of  $\Delta\rho$ and the $Z b_L \bar{b}_L$ coupling in the full LW theory. Our effective field
theory results, Eq.~(\ref{eq:rholeading}) and Eq.~(\ref{eq:Zbbleading}), provide a check of these full 
calculations, since the leading logarithmic contributions have to match. More generally, below we compute the top-sector one loop contribution to all Barbieri {\em et. al.} \cite{Barbieri:2004qk,Chivukula:2004af} electroweak parameters, and provide lower bounds on $M_q$ and $M_t$ from comparison with experiment.

\section{Constraints from Post-LEP Parameters}\label{sec:parameters}

In the language of Barbieri {\em et. al.} \cite{Barbieri:2004qk,Chivukula:2004af},  the observables $\hat{S}$, $\hat{T}$, $Y$, and $W$ parametrize the flavor-universal deviations from the SM at low energies. We now analyze the tree-level and the fermionic one-loop contibutions to these parameters, and use them to obtain constraints on the masses of the LW states.

\subsection{Tree-Level Contributions} 
At tree-level it is straightforward to read from Eq.~(\ref{eq:gaugelagrg}) the vacuum polarization amplitudes:
\begin{eqnarray}
\Pi_{\hat{W}^+ \hat{W}^-}(q^2) &=& \frac{q^2}{g_2^2} -\frac{(q^2)^2}{g_2^2\ M_2^2}-\frac{v^2}{4} \ , \nonumber \\
\Pi_{\hat{W}^3 \hat{W}^3}(q^2) &=& \frac{q^2}{g_2^2} -\frac{(q^2)^2}{g_2^2\ M_2^2}-\frac{v^2}{4} \ , \nonumber \\
\Pi_{\hat{W}^3 \hat{B}}(q^2) &=& \frac{v^2}{4} \ , \nonumber \\
\Pi_{\hat{B} \hat{B}}(q^2) &=& \frac{q^2}{g_1^2} -\frac{(q^2)^2}{g_1^2\ M_1^2}-\frac{v^2}{4} \ .
\label{eq:twopoint}
\end{eqnarray}
Following \cite{Barbieri:2004qk},  we see that there is no tree-level correction to the Fermi constant
\begin{eqnarray}
\frac{1}{\sqrt{2}G_F} = - 4 \Pi_{\hat{W}^+ \hat{W}^-}(0) = v^2 \ .
\label{eq:GF}
\end{eqnarray}
Barbieri {\em et. al.} define the electroweak gauge couplings
\begin{eqnarray}
\frac{1}{g^2} \equiv \Pi^\prime_{\hat{W}^+ \hat{W}^-}(0) \ , \label{eq:g}\\
\frac{1}{g^{\prime 2}} \equiv \Pi^\prime_{\hat{B} \hat{B}}(0) \ , \label{eq:ggprime} 
\end{eqnarray}
which in the LW SM gives $g'= g_1$ and $g = g_2$. We then compute the tree-level electroweak parameters~\cite{Underwood:2008cr},
\begin{eqnarray}
\hat{S} &\equiv & g^2\ \Pi^\prime_{\hat{W}^3 \hat{B}}(0) = 0 \ ,  \label{eq:treelevelShat}\\
\hat{T} &\equiv & g^2\left[\Pi_{\hat{W}^3 \hat{W}^3}(0)-\Pi_{\hat{W}^+ \hat{W}^-}(0)\right] = 0 \ , \\
Y &\equiv & \frac{1}{2}g^{\prime 2}m_W^2\ \Pi^{\prime\prime}_{\hat{B} \hat{B}}(0)=-\frac{m_W^2}{M^{2}_1} \ , \label{eq:treelevelY} \\
W &\equiv & \frac{1}{2}g^2m_W^2\ \Pi^{\prime\prime}_{\hat{W}^3 \hat{W}^3}(0)=-\frac{m_W^2}{M_2^2} \label{eq:treelevelW} \ ,
\end{eqnarray}
where in each equation the first equality is the definition of the corresponding post-LEP parameter \cite{Barbieri:2004qk}.

\subsection{Fermionic One-Loop Contributions}

\begin{figure}[!t]
\includegraphics[width=0.6\textwidth]{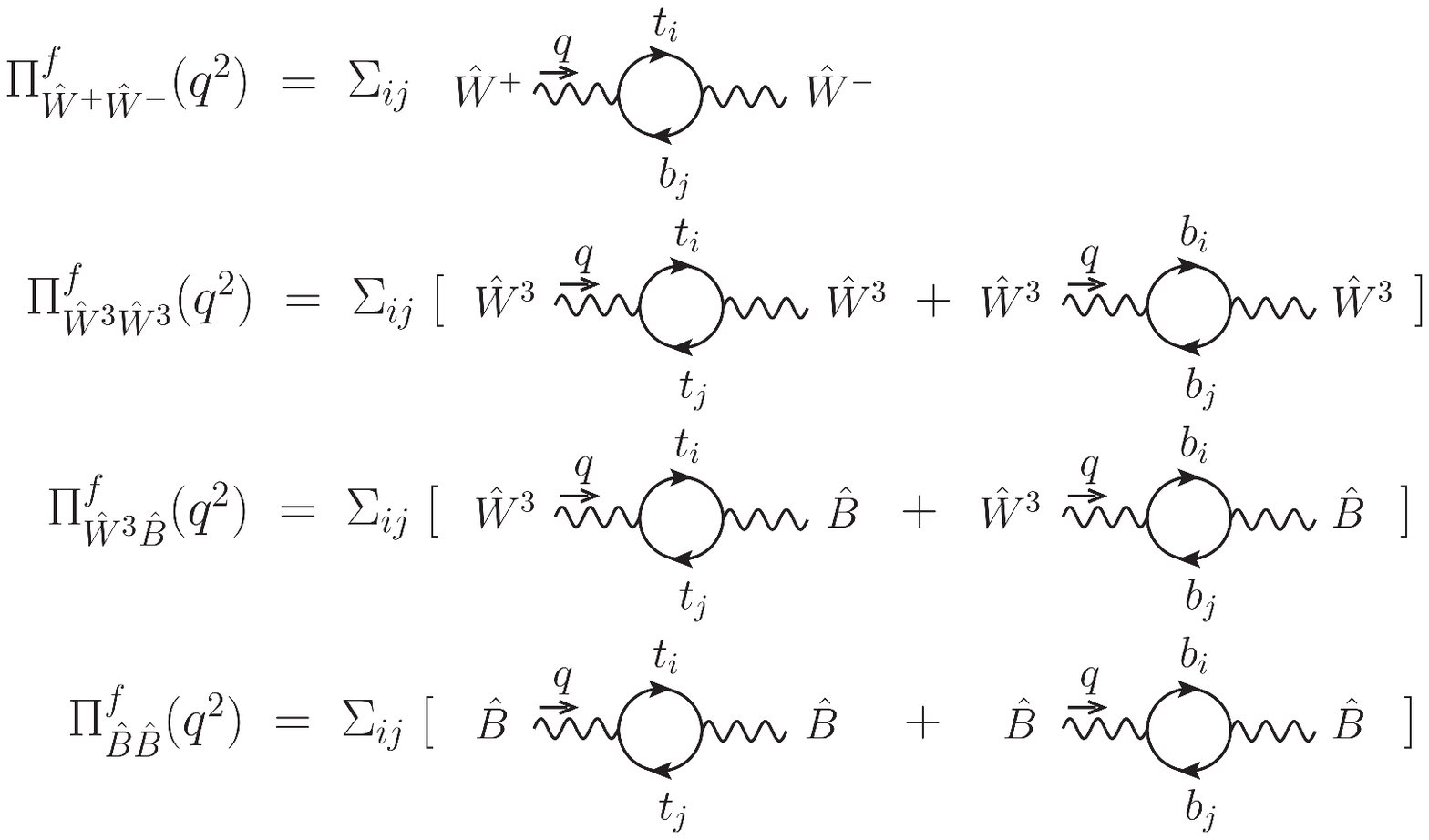}
\caption{Dominant vacuum polarization amplitudes for the LW SM gauge fields. These include the ordinary ($t_0$ and $b_0$) and the LW third generation quarks ($t_1$, $t_2$, $b_1$, and $b_2$) in the loop. These amplitudes contribute to the two-point functions of Eq.~(\ref{eq:twopoint}).}
\label{fig:loops}
\end{figure}

The gauge current correlators receive important loop corrections from the top-bottom sector, through the diagrams shown in Fig.~{\ref{fig:loops}}. These vacuum polarization amplitudes contain two infinities, which are absorbed in the definitions of $g$ and $g^\prime$ given in Eqs.~(\ref{eq:g}) and (\ref{eq:ggprime}), respectively. 
As a consequence the noncanonical normalization adopted in Eq.~(\ref{eq:gaugelagrg}) forces us to define renormalized LW gauge masses. A convenient scheme consists of defining $M$ and $M^\prime$ by
\begin{equation}
 -\,\frac{2}{g^2 M^2} \equiv \Pi''_{\hat{W}^+ \hat{W}^-}(0) \qquad \qquad -\,\frac{2}{g^2 M'^2} \equiv \Pi''_{\hat{B}\hat{B}}(0)~,
\label{eq:newdef}
\end{equation}
which simplify the one-loop calculations below. At tree-level, from 
Eq.~(\ref{eq:twopoint}), we see that $M=M_2$ and $M^\prime=M_1$, and both are related
to the masses of the LW partners of the gauge-bosons. Due to the power-counting
properties of LW theories, after the usual\footnote{Notice that the vacuum polarization diagrams involving only one LW fermion carry an overall negative sign. In fact this happens to make all zero-derivative functions, at $q^2=0$, finite. For this reason there is actually one less infinity compared to the ordinary SM, and the bare $v$ is finite \protect\cite{Grinstein:2007mp}.} coupling-constant and mass renormalizations,
all physical quantities remain finite \cite{Grinstein:2007mp}. Hence $M$ and $M'$ remain
finite at one-loop (and beyond). However, since they are defined by the zero-momentum properties
of the gauge-boson two-point functions their values only approximately equal the masses
of the LW partners of the gauge-bosons.  This suffices for our purposes, since we are interested in low-energy observables; if we were studying quantities measured at higher energies, we would want to define $M$ and $M'$ based on propagators renormalized at high $q^2$ instead.

The propagators in the loops of Fig.~\ref{fig:loops} correspond to mass eigenstates, where the masses are obtained by diagonalizing the mass matrices by means of symplectic rotations: in this way the LW fields maintain the ``wrong-sign'' kinetic and mass terms. A perturbative diagonalization in $v^2/M_q^2$ and $v^2/M_t^2$~\cite{Alvarez:2008ks} requires considering two different scenarios: nondegenerate LW masses, $|M_q^2-M_t^2|\sim M_q^2$, and (near) degenerate LW masses, $|M_q^2-M_t^2|\ll M_q^2$. For nondegenerate LW top quarks the contributions to the electroweak parameters are quite lengthy. To leading order we obtain
\begin{eqnarray}
\hat{S} & = & -\frac{g^2 m_t^2}{48\pi^2 M_q^2}\Bigg[
\left(2+\frac{1}{r_t^2}\right)\log\frac{M_q^2}{m_t^2}
+\frac{1-3r_t^2+6r_t^4-r_t^6+3r_t^8}{r_t^2(1-r_t^2)^5}\log r_t^2
-\frac{5-17r_t^2+4r_t^4+12r_t^6-23r_t^8+7r_t^{10}}{2r_t^2(1-r_t^2)^4}\Bigg] \ , \nonumber \\
\hat{T} & = & -\frac{3 g^2 m_t^4}{32\pi^2 m_W^2 M_q^2}\Bigg[
\left(2+\frac{1}{r_t^2}\right)\log\frac{M_q^2}{m_t^2}
+\frac{1-3r_t^2+6r_t^6}{r_t^2(1-r_t^2)^5}\log r_t^2
-\frac{9-12r_t^2-21r_t^4+46r_t^6-68r_t^8+22r_t^{10}}{6r_t^2(1-r_t^2)^4}\Bigg] \ , \nonumber \\
Y & = & -\frac{m_W^2}{M^{\prime 2}} \ , \nonumber \\
W & = & -\frac{m_W^2}{M^2}+\frac{g^2 m_W^2}{640\pi^2 M_q^2}\Bigg[-7+\frac{3}{r_b^2}-\frac{9}{r_t^2}\Bigg] \ ,
\label{eq:nondeg}
\end{eqnarray}
where
\begin{eqnarray}
r_t\equiv M_t/M_q \ , \quad r_b \equiv M_b/M_q \ .
\end{eqnarray}
The electroweak parameters in the (near) degenerate case cannot simply be obtained by taking the $r_t, r_b \to 1$ limit in Eqs.~(\ref{eq:nondeg}), since the corresponding expressions diverge. Instead we must diagonalize the mass matrices perturbatively in $1/M_q^2$ (or $1/M_t^2$) and $|M_q^2-M_t^2|/M_q^2$, and then compute the electroweak parameters. For exact degeneracy, $M_q=M_t$, this gives
\begin{eqnarray}
\hat{S} & = & -\frac{g^2 m_t^2}{16\pi^2 M_q^2}\Bigg[\log\frac{M_q^2}{m_t^2}-\frac{12}{5}\Bigg] \nonumber \\
\hat{T} & = & -\frac{3 g^2 m_t^4}{32\pi^2 m_W^2 M_q^2}\Bigg[3\log\frac{M_q^2}{m_t^2}-\frac{141}{20}\Bigg] \nonumber \\
Y & = & -\frac{m_W^2}{M^{\prime 2}} \nonumber \\
W & = & -\frac{m_W^2}{M^2}-\frac{7 g^2 m_W^2}{640\pi^2 M_q^2} \ .
\label{eq:deg}
\end{eqnarray}

Note that the absence of fermionic one-loop corrections to the tree-level value of $Y$ is a direct consequence of the second definition in Eq.~(\ref{eq:newdef}): a different scheme choice would lead to an additional contribution. In the same way, changing the definition of $M$ would lead to a different fermionic one-loop expression for $W$; in any case, the second term\footnote{Note that the first definition in Eq.~(\protect\ref{eq:newdef}) pertains to
$\Pi''_{\hat{W}^+\hat{W}^-}$, whereas $W$ is defined in terms of $\Pi''_{\hat{W}_3 \hat{W}_3}$.}
 in $W$ is numerically very small and can be neglected. We therefore conclude that the leading contributions to $Y$ and $W$ are those arising from the LW gauge-sector at tree-level, Eqs.~(\ref{eq:treelevelY}, \ref{eq:treelevelW}).

Since the tree-level values of $\hat{S}$ and $\hat{T}$ vanish, the leading LW contributions to both $\hat{S}$ and $\hat{T}$ arise from the top-quark sector at one loop. In the case of $\hat{T}$ this is not surprising since the dominant locus of isospin violation in the model is the splitting between the top and bottom quark masses. Because $\hat{T}$ is the same as $\Delta\rho$~\cite{Chivukula:2004af}, we may
compare the leading logarithmic correction in Eq.~(\ref{eq:nondeg}) with  the result obtained in the effective theory, Eq.~(\ref{eq:rholeading}); we see that they agree. In the case of $\hat{S}$,
the situation is more subtle. The LW gauge-eigenstate fermion partners, being massive, 
are not chiral and therefore, in the absence of electroweak contributions to the masses 
which mix them with the light chiral gauge-eigenstates, their contribution to $\hat{S}$ vanishes.
Hence the dominant LW contributions to $\hat{S}$ also arise predominantly from the
top-sector of the theory.

Therefore, at tree-level plus one fermion loop we obtain a very simple conclusion: the fermion sector contributes to $\hat{S}$ and $\hat{T}$ only, while the gauge sector contributes to $Y$ and $W$ only.  It is true that when gauge loops are included, there will be additional contributions.  However, the gauge loop contributions will be sub-dominant compared to the quantities we have already calculated; the only potential exception is $\hat{S}$, for which all of the one-loop contributions are too small to be experimentally relevant.  Thus, our existing results suffice for extracting constraints from the experimental data.

\subsection{Comparison with Data}

\begin{figure}[!t]
\includegraphics[width=3in]{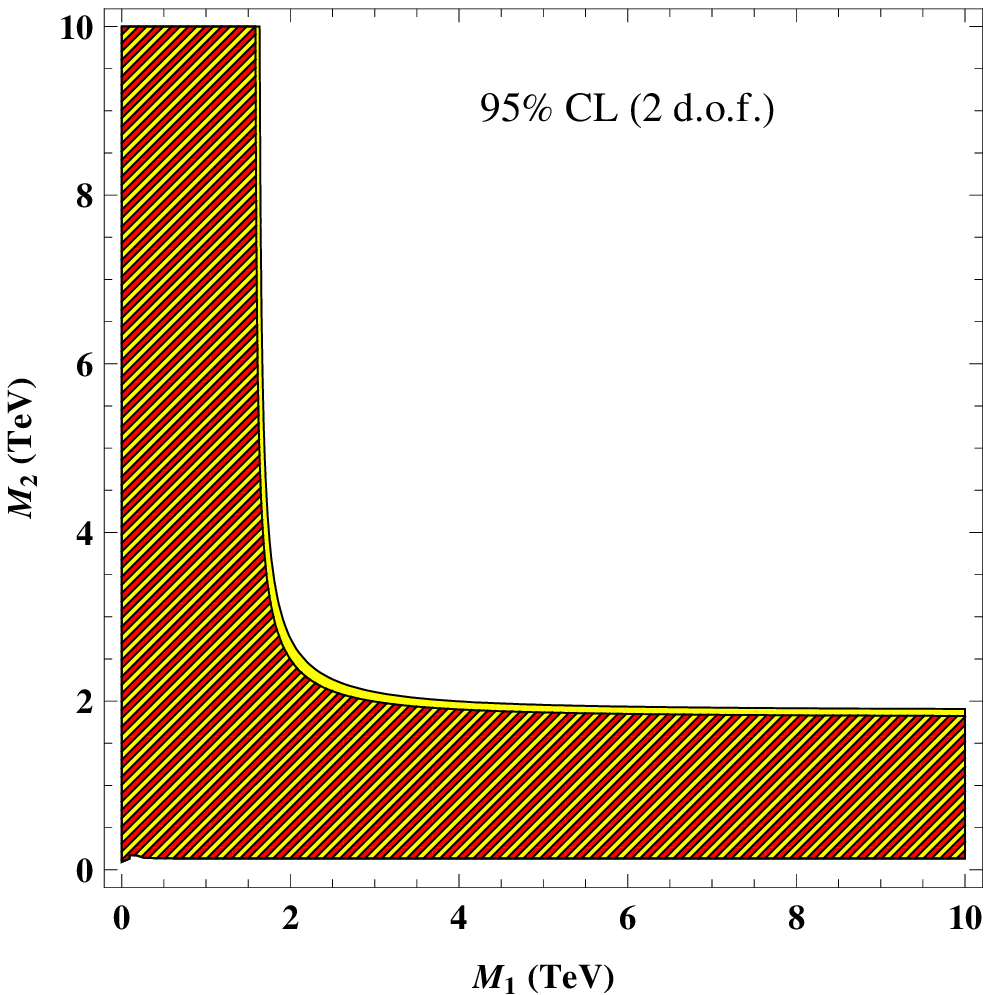}
\includegraphics[width=3in]{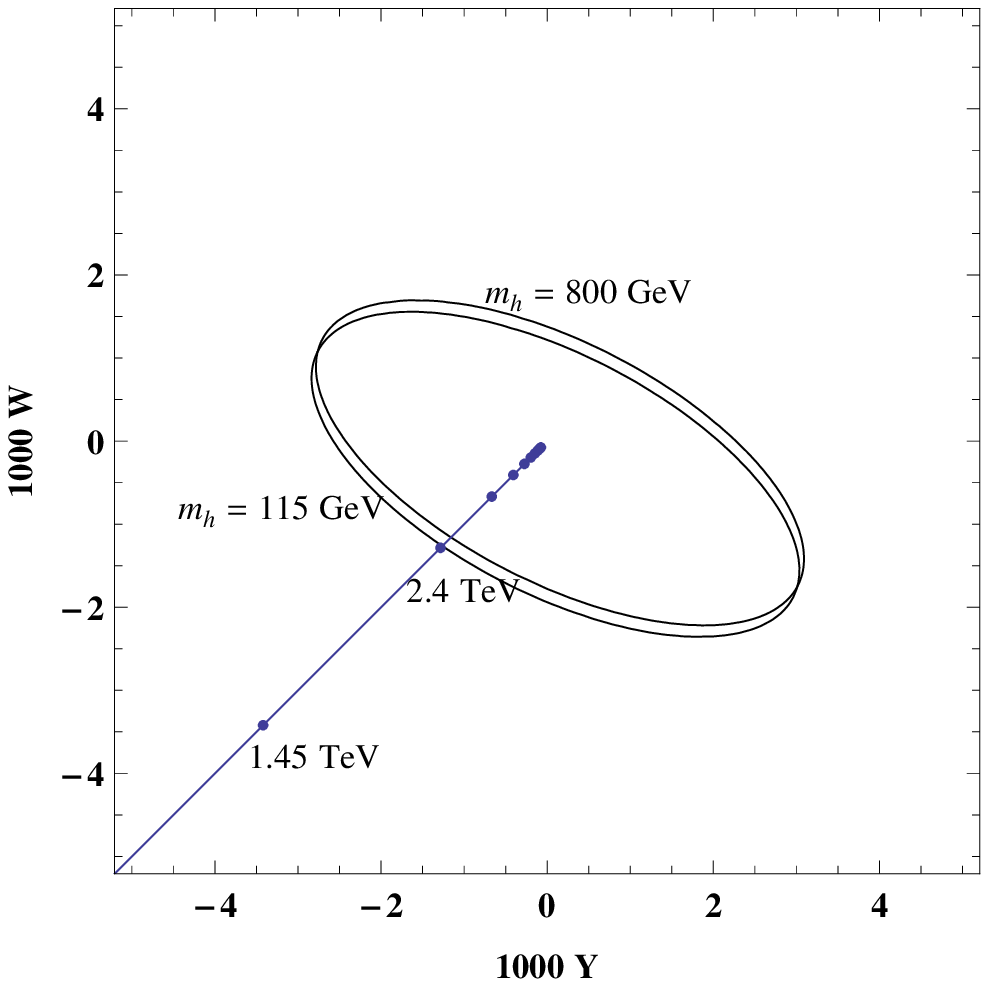}
\caption{Left: Exclusion plot for the LW gauge-field masses $M_2$ and $M_1$. These bounds are due to the constraints on $Y$ and $W$, as shown by Eq.~(\ref{eq:treelevelY}) and Eq.~(\ref{eq:treelevelW}). For a light Higgs ($m_h=115$ GeV) the striped region to the left of both curves is excluded. For a heavy Higgs ($m_h=800$ GeV) the additional yellow strip between the curves is excluded as well. Right: 95\% C.L. ellipses in the ($Y$,$W$) plane, and the LW prediction for degenerate masses, $M_1=M_2$. The parametric plot is for  $0.5\ {\rm TeV} < M_1 = M_2 < 10\ {\rm TeV}$ and the dots are equally spaced in mass. The lower bound on $M_1 = M_2$ is approximately 2.4 TeV for a light Higgs.}
\label{fig:gauge}
\end{figure}

We begin with constraints on the masses of the LW partners of the gauge bosons.  The previous subsection found that  the only post-LEP parameters affected by the LW gauge boson masses are $W$ and $Y$, and also that the tree-level expressions for $W$ and $Y$, Eqs.~(\ref{eq:treelevelY}, \ref{eq:treelevelW}), suffice for comparison with data.   The experimental constraints on $Y$ and $W$ are rather tight, and almost independent of the value of the Higgs mass~\cite{Barbieri:2004qk}. These translate into the 95\% CL lower bounds on $M_2$ and $M_1$ shown\footnote{These bounds are derived using the errors and correlation matrix given in Ref.~\cite{Barbieri:2004qk}} in Fig.~\ref{fig:gauge}. The left plot shows the bounds for arbitrary values of $M_1$ and $M_2$: for $m_h=115$ GeV the striped region is excluded, while for $m_h=800$ GeV the additional narrow yellow region is excluded as well. The right plot shows the 95\% C.L. ellipses in the $(Y,W)$ plane from the global fit to data~\cite{Barbieri:2004qk}, for $m_h=115$ and $m_h=800$, as well as the LW prediction for degenerate LW masses, $M_1=M_2$. All this is in agreement with the results of Ref.~\cite{Underwood:2008cr} and gives the constraints $M_1$, $M_2\ \gtrsim 2.4$ TeV.

Next, we seek constraints on the masses of the LW partners of the top quark.  The previous subsection found that the post-LEP parameters sensitive to the LW fermion masses are $\hat{S}$ and $\hat{T}$, which do not depend on the LW gauge masses at the one-loop level. We should also note that, for a light Higgs, the LW prediction of $\hat{S}$ is very close to its central value, $\hat{S}\simeq0$. Furthermore from the global fit to the experimental data in Ref.~\cite{Barbieri:2004qk}, we conclude that $\hat{T}$ is only mildly correlated to $Y$ and $W$, the parameters that are most sensitive to the LW gauge boson masses in the LW SM. This confirms that the bounds on the LW fermions should be essentially independent of the LW gauge masses, and should come almost entirely from $\hat{T}$. 

\begin{figure}
\begin{center}
\includegraphics[width=4in]{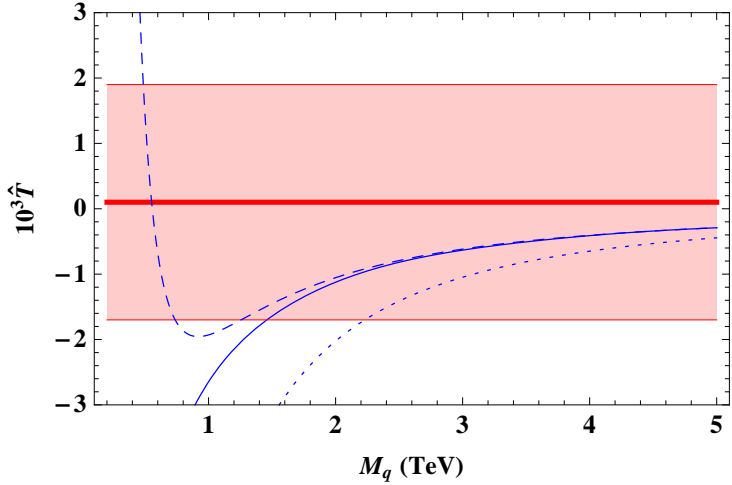}
\caption{$\hat{T}$ as a function of $M_q$ in the degenerate case, $M_q=M_t$. 
The experimental mean value for $\hat{T}$ is shown by the thick red line, along with
the $\pm 2 \sigma$ allowed region. Also shown are the all-order (in $v^2/M_q^2$) LW prediction (solid blue curve), the leading order LW prediction, Eq.~(\protect\ref{eq:deg}) (dashed blue curve), and the leading-log curve, Eq.~(\protect\ref{eq:rholeading})  (dotted blue curve), as functions of $M_q$, in the degenerate case.  Note that the leading-order prediction is not valid below $M_q \sim 1$ TeV. (See text for details.)}
\label{fig:That}
\end{center}
\end{figure}

In Fig.~\ref{fig:That} we show the experimental mean value for $\hat{T}$ (red thick line), the $\pm 2 \sigma$ allowed region, the all-order (in $v^2/M_q^2$) LW prediction (solid blue curve), the leading order LW prediction from Eq.~(\ref{eq:deg}) (dashed blue curve), and the leading-log approximation (dotted blue curve), as functions of $M_q$, in the degenerate case. This figure reveals the bound $M_q=M_t\gtrsim$ 1.6 TeV   on the LW fermion masses in the degenerate case. Note that although Eq.~(\ref{eq:deg}) appears to predict a positive $\hat{T}$ for small $M_q$ (dashed blue curve), the complete numerical evaluation (solid blue curve) shows that $\hat{T}$ is always negative, as Fig.~\ref{fig:That} shows explicitly; below $M_q=1$ TeV the perturbative diagonalization of the mass matrix is no longer valid, rendering the leading-order LW prediction unreliable in that mass regime.

\begin{figure}[!t]
\includegraphics[width=3in]{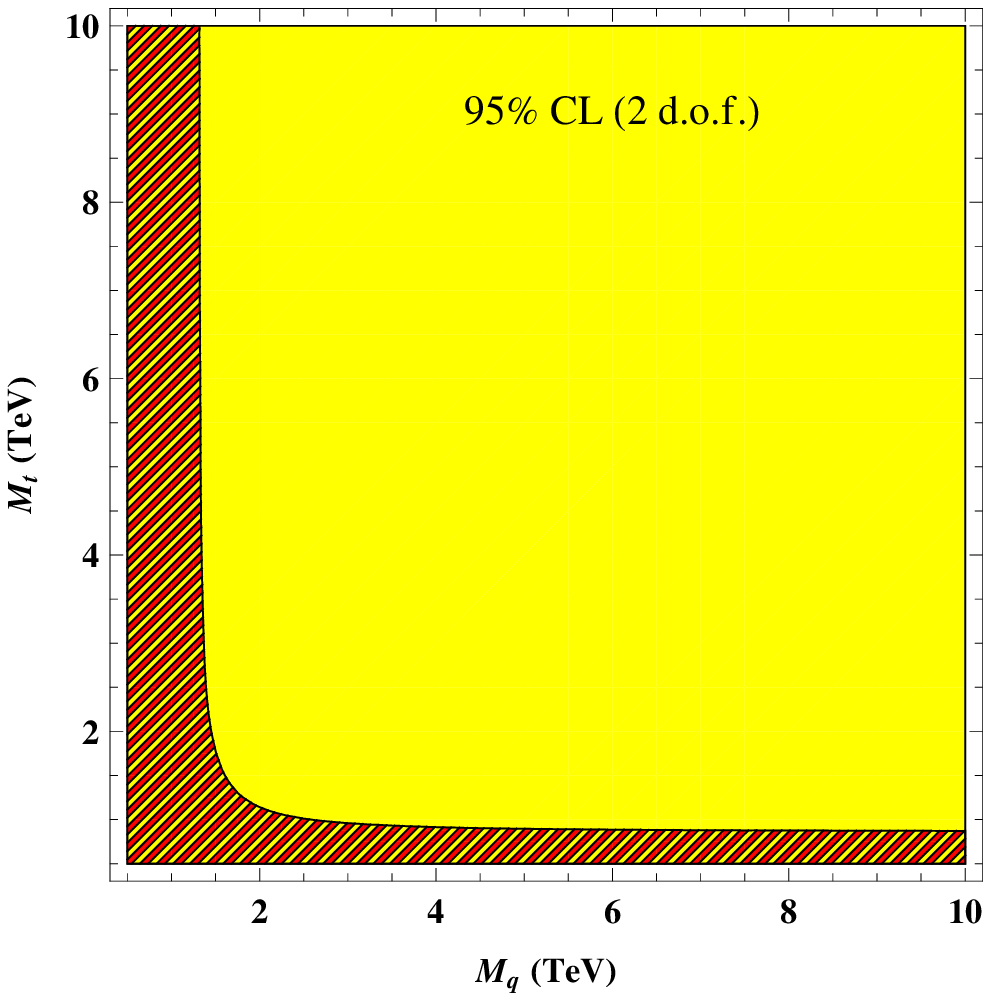}
\includegraphics[width=3in]{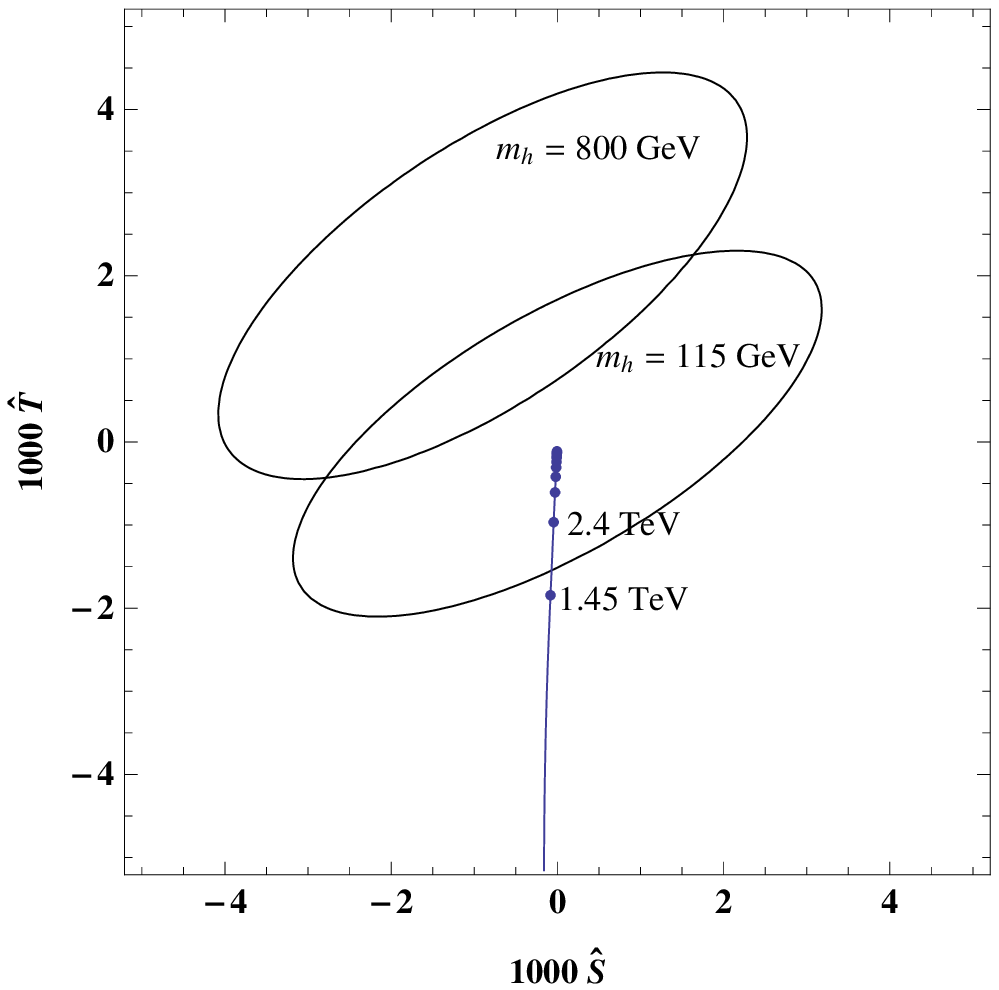}
\caption{Left: 95\% C.L. exclusion plots for the LW fermion masses $M_q$ and $M_t$. These bounds come almost entirely from the experimental constraints on $\hat{T}$. For a light Higgs the striped region to the left of the curve is excluded, while a heavy Higgs is completely excluded. Right: 95\% C.L. ellipses in the ($\hat{S}$,$\hat{T}$) plane, and the LW prediction for degenerate masses, $M_q=M_t$. The parametric plot is for  $0.5\ {\rm TeV} < M_q < 10\ {\rm TeV}$ and the dots are equally spaced in mass. The lower bound on $M_q$ is approximately 1.5 TeV for a light Higgs.}
\label{fig:fermionbounds}
\end{figure}

If we relax the requirement of degenerate LW fermion masses, we obtain the 95\% C.L. bounds on $M_q$ and $M_t$ shown in Fig.~\ref{fig:fermionbounds} (left).  For a light Higgs the striped region in Fig.~\ref{fig:fermionbounds} (left) is excluded, while for a heavy Higgs the whole (yellow) region is excluded.  Note from Figs.~\ref{fig:gauge} and \ref{fig:fermionbounds} (left) that the mildest constraints on the LW masses are obtained in the fully degenerate case, $M=M^\prime$ and $M_q=M_t$. 

Returning to the case of degenerate LW fermion masses, we show in Fig.~\ref{fig:fermionbounds} (right) the values of $\hat{S}$ and $\hat{T}$ as a function of $M_q=M_t$  for $0.5\ {\rm TeV} < M_q < 10\ {\rm TeV}$; the dots representing different values of $M_q$ are placed at regular intervals. The 95\% C.L. ellipses from the global fit to the data~\cite{Barbieri:2004qk} confirm the constraint $M_q\gtrsim 1.5$ TeV for a light Higgs, while a heavy Higgs scenario is disfavored for any LW fermion mass.   In fact for a heavy Higgs the $\hat{T}$ parameter is expected to be positive, while the LW SM predicts a negative $\hat{T}$. This is a direct consequence of the negative sign in the LW fermion propagators, which results in an overall negative sign from the (dominant) diagrams involving a single LW fermion in the loop.

Our results disagree with those of  \cite{Alvarez:2008ks,Carone:2008bs} in two ways: their bounds on the LW fermion masses appear more stringent for a light Higgs and their limits appear to depend on the masses of the LW gauge boson partners.  The disagreement arises because their study of one-loop electroweak corrections in the LW SM assumes the corrections  to be purely oblique, and derives constraints by comparing the Peskin-Takeuchi $S$ and $T$~\cite{Peskin:1991sw} parameters to data. However, as clearly discussed in Ref.~\cite{Underwood:2008cr}, and confirmed above in Eqs.~(\ref{eq:treelevelY}) and (\ref{eq:treelevelW}), the LW SM features large non-oblique corrections, in the form of  non-zero values for $Y$ and $W$ at tree-level.  Hence, one must use the Barbieri {\em et. al.}  parameters to compare the LW SM with experiment, as we have done.

\section{Constraints from the $Z b_L \bar{b}_L$ Coupling}\label{sec:Zbb}

\begin{figure}[!t]
\includegraphics[width=3.5in,height=2.0in]{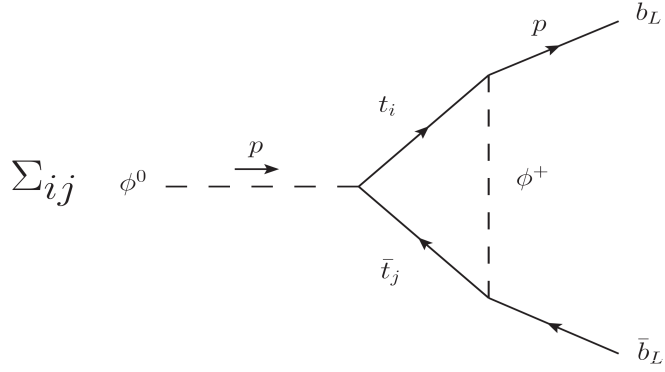}
\caption{Diagram giving the largest contribution to the $Z b_L \bar{b}_L$ coupling. The latter is related through the Ward identity to the $\phi^0 b_L \bar{b}_L$ coupling -- where $\phi^0$ is the Goldstone boson eaten by the $Z$ boson. The top quarks running in the loop are both ordinary and LW.}
\label{fig:Zbb}
\end{figure}

The leading contribution to the $Z b_L \bar{b}_L$ coupling (in the gauge coupling expansion) can be obtained in the gaugeless limit from the  $\phi^0 b_L \bar{b}_L$ coupling~\cite{Barbieri:1992nz,Barbieri:1992dq,Oliver:2002up,Abe:2009ni}, where $\phi^0$ is the Goldstone boson eaten by the $Z$. The loop diagram giving the largest correction involves the SM and LW top quarks\footnote{In the gaugeless limit of the LW SM, as in the SM itself, all external b-quark
wavefunction renormalization corrections are proportional to $y^2_b$ and
are therefore negligible. This should be contrasted with the situation
in the three-site Higgsless model [33].}
, and is shown in Fig.~\ref{fig:Zbb}. A detailed computation of the loop integral, valid for arbitrary models with heavy replicas of the top quark, is given in the appendix. At zero external momentum the amplitude corresponding to the diagram has the form
\begin{eqnarray}
i\ M = - A\ \slashed{p} P_L \ ,
\label{eq:MZbb}
\end{eqnarray}
where $P_L\equiv (1-\gamma^5)/2$ is the left-handed projector, $p$ is the incoming $\phi^0$ momentum, and the external fermion wavefunctions have been omitted. Then to leading order in $g$ the correction to the $Z b_L \bar{b}_L$ coupling is~\cite{Barbieri:1992nz,Barbieri:1992dq,Oliver:2002up,Abe:2009ni}
\begin{eqnarray}
\delta g_L^{b\bar{b}} = \frac{v}{2} A \ . 
\label{eq:generalZBB}
\end{eqnarray}
Expanding the amplitude in powers of $m_t^2/M_q^2$ we obtain
\begin{eqnarray}
(\delta g_L^{b\bar{b}})_{\rm LW}=-\frac{m_t^4}{32\pi^2 v^2 M_q^2}\Bigg[\left(\frac{4}{r_t^2}+1\right)\log\frac{M_q^2}{m_t^2}
+\frac{4-11 r_t^2+9 r_t^4}{r_t^2(1-r_t^2)^3}\log r_t^2 -\frac{6-10 r_t^2 + 2r_t^4}{r_t^2(1-r_t^2)^2}\Bigg] 
\label{eq:Zbbnondeg}
\end{eqnarray}
for nondegenerate LW fermion masses, and
\begin{eqnarray}
(\delta g_L^{b\bar{b}})_{\rm LW}=-\frac{m_t^4}{32\pi^2 v^2 M_q^2}\Bigg[5\log\frac{M_q^2}{m_t^2}-\frac{49}{6}\Bigg]
\label{eq:Zbbdeg}
\end{eqnarray}
for degenerate LW masses. Both of these expressions agree with the dominant contribution found in the effective theory, Eq.~(\ref{eq:Zbbleading}).  

\begin{figure}[!t]
\includegraphics[width=4in]{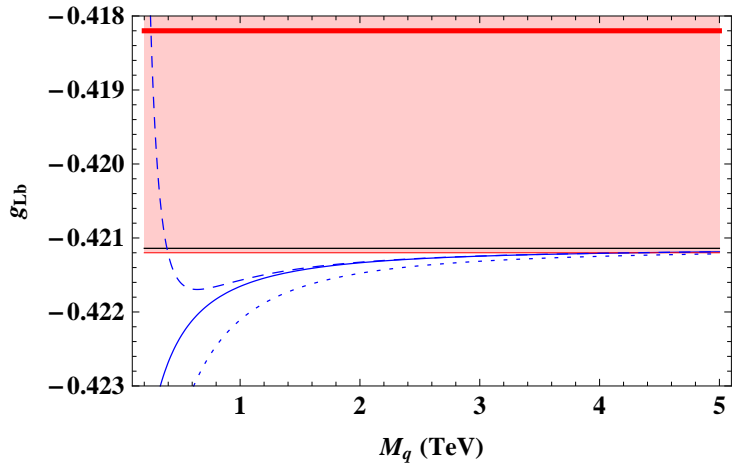}
\caption{Constraints from the $Z b_L \bar{b}_L$ coupling. This graph features the experimental mean value (red thick horizontal line), the 2$\sigma$ allowed region below the mean value, the SM prediction (solid horizontal black line), the all-order (in $v^2/M_q^2$) LW prediction (solid blue curve), the leading order LW prediction from Eq.~(\ref{eq:Zbbdeg}) (dashed blue curve), and the leading-log approximation (dotted blue curve), as functions of $M_q$, in the degenerate case.}
\label{fig:Zbbexclusion}
\end{figure}

The experimental value of $g_L^{b\bar{b}}$ is derived from measurements of $R_b$, the ratio of the $Z\to b \bar{b}$ width to the width of the hadronic decays, and $A_b$, the forward-backward asymmetry for $Z$ decays into $b\bar{b}$~\cite{:2005ema}:
\begin{eqnarray}
(g_L^{b\bar{b}})_{\rm exp}=-0.4182\pm 0.0015 \ .
\end{eqnarray}
The SM value was computed using ZFITTER~\cite{Bardin:1999yd,Arbuzov:2005ma} in Ref.~\cite{SekharChivukula:2009if}, leading to 
\begin{eqnarray}
(g_L^{b\bar{b}})_{\rm SM}=-\frac{1}{2}  + \frac{1}{3}\sin ^2 \theta_W 
+ (\delta g_L^{b\bar{b}})_{\rm SM}=-0.42114 \ ,
\end{eqnarray}
while the LW prediction is given by Eq.~(\ref{eq:Zbbnondeg}) and Eq.~(\ref{eq:Zbbdeg}). In Fig.~\ref{fig:Zbbexclusion} we show the experimental mean value (red thick horizontal line), the 2$\sigma$ allowed region below the mean value, the SM prediction (solid horizontal black line), the all-order (in $v^2/M_q^2$) LW prediction (solid blue curve), the leading order LW prediction from Eq.~(\ref{eq:Zbbdeg}) (dashed blue curve), and the leading-log approximation (dotted blue curve), as functions of $M_q$, in the degenerate case. Note that the dashed curve and Eq.~(\ref{eq:Zbbdeg}) are not reliable for $M_q\lesssim 1$ TeV, because the perturbative diagonalization of the mass matrix is not valid in that mass regime.

As anticipated by the effective field theory calculation, the LW correction is always negative: this is essentially due to the negative sign in front of the dominant nonstansdard triangle diagrams with one LW top and one SM top. It is large (for small values of $M_q$) because of the explicit breaking of custodial isospin symmetry. Since the SM value is already 1.96$\sigma$ below the experimental mean value, this correction goes in the direction opposite to what is needed. Agreement at the 2$\sigma$ level requires $M_q\gtrsim 4$ TeV; at 2.5$\sigma$ this bound relaxes to $M_q\gtrsim$ 700 GeV.

\section{Conclusions}\label{sec:conclusions}

There is significant tension between naturalness and isospin violation in the Lee-Wick Standard Model (LW SM).  While corrections to the Higgs mass are smallest when the LW partners of the gauge bosons and fermions are light, isospin violation that must be present in the top sector to account for the large splitting between $m_t$ and $m_b$ tends to constrain the LW partners to have masses over a TeV.  We have performed an effective field theory analysis of the corrections to $\hat{T}$ and the $Z b_L \bar{b}_L$ coupling in the LW SM and used it to confirm our full calculation of the LW effects on $\hat{S}$, $\hat{T}$, $W$, $Y$ and $(g_L^{b\bar{b}})$ including tree-level and fermionic one-loop contributions. The post-LEP parameters yield a simple picture in the LW SM:  the gauge sector contributes to $Y$ and $W$ only, with leading contributions arising at tree-level, while the fermion sector contributes to $\hat{S}$ and $\hat{T}$ only, with leading corrections arising at one loop.  

In agreement with \cite{Underwood:2008cr}, we find that experimental limits on $W$ and $Y$ jointly constrain the masses of the LW gauge bosons to satisfy  $M_1, M_2  \gtrsim 2.3$ TeV at 95\% CL, with relatively little sensitivity to the Higgs mass.

We also conclude that the experimental limits on $\hat{T}$ require the masses of the LW fermions to satisfy $M_q, M_t  \gtrsim 1.5$ TeV at 95\% CL if the Higgs mass is light and tend to exclude the LW SM for any LW fermion masses if the Higgs mass is heavy.   This is because a model containing a heavy Higgs can be rendered consistent with the data only if some other sector of the model provides a large positive correction to $\hat{T}$.  However, in the LW SM, the fermionic loops that provide the dominant contribution to $\hat{T}$ always make $\hat{T}$ more negative, due to the negative sign in the LW fermion propagators.  The LW fermions simply cannot compensate for the presence of a heavy Higgs.  Our results differ from those in Refs.~\cite{Alvarez:2008ks, Carone:2008bs} because their analysis incorrectly assumes that the electroweak corrections due to LW states are purely oblique.  As explained in Ref.~\cite{Underwood:2008cr} the LW states actually induce important non-oblique corrections, and one must therefore use the Barbieri {\em et. al.}  ~\cite{Barbieri:2004qk,Chivukula:2004af} post-LEP parameters to compare the LW SM with experiment, as we have done.

Weak isospin violation in the top-quark sector also manifests itself through corrections to the $Zb_L \bar{b}_L$ coupling.  The SM prediction for $(g_L^{b\bar{b}})$ lies at the lower end of the range allowed by experiment at 95\% CL, so that new physics making negative contributions to the value of $(g_L^{b\bar{b}})$ would decrease the agreement with the data.   As in the case of $\hat{T}$, however, we find that the negative sign in the LW fermion propagators translates into a negative contribution to $(g_L^{b\bar{b}})$; the lighter the LW fermions, the greater the disagreement between prediction and data.  We find that the $Z b_L \bar{b}_L$ coupling places a lower bound of 4 TeV on the LW fermion masses at 95\% CL.

\appendix*

\section{Evaluation of the $\phi^0\to b_L\bar{b}_L$ Amplitude}

The triangle diagram of Fig.~\ref{fig:Zbb} can be easily evaluated once the mass matrix has been diagonalized and the Yukawa couplings have been computed. For an arbitrary theory with heavy replicas of the third generation quarks, and neglecting the bottom Yukawa sector, the interactions with the pions eaten by the $W$ and $Z$ boson read 
\begin{eqnarray}
-i\ \frac{y_t}{\sqrt{2}}\ \phi^0 \left[\alpha_{ij}\ \bar{t_i} P_R t_j - \alpha_{ji}\ \bar{t_i} P_L t_j\right]
-i\ y_t\  \beta_{ij} \left[\phi^- \bar{b}_i P_R t_j - \phi^+ \bar{t}_j P_L b_i\right] \ ,
\end{eqnarray}
where $t_1$ and $b_1$ are the SM top and bottom, respectively, the remaining ones are heavy replicas, and where repeated indices are summed. From this expression one may extract the Feynman rules. Shifting the momentum of the $\bar{b}_L$ to zero, and omitting the external fermion wavefunctions, the amplitude reads
\begin{eqnarray}
i\ M = \sum_{i,j}(-)^{N_{ij}} \int\frac{d^4 k}{(2\pi)^4}  (y_t \beta_{1i} P_R) \frac{i(\slashed{k}+\slashed{p}+m_{t_i})}{(k+p)^2-m_{t_i}^2+i\epsilon}\frac{y_t}{\sqrt{2}}(\alpha_{ij}P_R-\alpha_{ji}P_L)
\frac{i(\slashed{k}+m_{t_j})}{k^2-m_{t_j}^2+i\epsilon}(-y_t \beta_{1j} P_L)\frac{i}{k^2+i\epsilon} \ , \qquad
\end{eqnarray}
where $N_{ij}$ is the number of LW fermions in the $i,j$ couple. Combining the denominators into a single one, and shifting the loop momentum in the usual way, leads to
\begin{eqnarray}
i\ M &=& -\frac{i y_t^3}{\sqrt{2}}\ \slashed{p} P_L \sum_{i,j}(-)^{N_{ij}} \beta_{1i}\beta_{1j}\alpha_{ji} m_{t_j}
\int_0^1 dx \int_0^{1-x} dy\ \int \frac{d^4 l}{(2\pi)^4} \frac{2(1-x)}{(l^2-\Delta)^3} \nonumber \\
&-& \frac{i y_t^3}{\sqrt{2}}\ \slashed{p} P_L \sum_{i,j}(-)^{N_{ij}} \beta_{1i}\beta_{1j}\alpha_{ij} m_{t_i}
\int_0^1 dx \int_0^{1-x} dy\ \int \frac{d^4 l}{(2\pi)^4} \frac{2x}{(l^2-\Delta)^3} \ ,
\end{eqnarray}
where
\begin{eqnarray}
\Delta \equiv -x(1-x)p^2+x m_{t_i}^2 + y m_{t_j}^2 \ .
\end{eqnarray}
Evaluating the integrals in the $p^2\to 0$ limit gives
\begin{eqnarray}
i\ M &=& -\frac{1}{16\pi^2}\frac{y_t^3}{\sqrt{2}}\ \slashed{p} P_L
\Bigg[\sum_i \frac{\beta_{1i}^2 \alpha_{ii}}{m_{t_i}} \nonumber \\
&+&\sum_{i \neq j}(-)^{N_{ij}} \beta_{1i}\beta_{1j}\alpha_{ji} m_{t_j}
\Bigg(-\frac{1}{m_{t_i}^2-m_{t_j}^2}
+\frac{1}{2}\frac{3m_{t_i}^2-m_{t_j}^2}{(m_{t_i}^2-m_{t_j}^2)^2}\log\frac{m_{t_i}^2}{m_{t_j}^2}\Bigg)\Bigg] \ .
\end{eqnarray}
Comparing this expression with Eq.~(\ref{eq:MZbb}) and Eq.~(\ref{eq:generalZBB}) gives
\begin{eqnarray}
\delta g_L^{b\bar{b}} = \frac{1}{16\pi^2}\frac{y_t^3 v}{2\sqrt{2}}
\Bigg[\sum_i \frac{\beta_{1i}^2 \alpha_{ii}}{m_{t_i}}
+\sum_{i \neq j}(-)^{N_{ij}} \beta_{1i}\beta_{1j}\alpha_{ji} m_{t_j}
\Bigg(-\frac{1}{m_{t_i}^2-m_{t_j}^2}
+\frac{1}{2}\frac{3m_{t_i}^2-m_{t_j}^2}{(m_{t_i}^2-m_{t_j}^2)^2}\log\frac{m_{t_i}^2}{m_{t_j}^2}\Bigg)\Bigg] \ 
\end{eqnarray}
to leading order in the weak gauge coupling.

\end{document}